\documentclass[preprint,showpacs,showkeys,preprintnumbers,amsmath,amssymb]{revtex4}
\usepackage{graphicx}
\usepackage{dcolumn}
\usepackage{bm}
\begin{document}
\preprint{APS/123-QED}
\title{Ensemble learning of linear perceptrons;\\Online learning theory}
\author{Kazuyuki Hara${}^{\dagger}$}
\email{hara@tokyo-tmct.ac.jp}
\author{Masato  Okada${}^{\ddagger}$}
\affiliation{
${}^{\dagger}$Department of Electronics and Information Engineering, Tokyo Metropolitan College of Technology, 
Higashi-oi 1-10-40, Shinagawa-ku, Tokyo, 140-0011 Japan\\
${}^{\ddagger}$Laboratory for Mathematical Neuroscience, RIKEN Brain Science Institute,
Hirosawa 2-1, Wakou-shi, Saitama, 351-0456 Japan
}
\date{\today}
\begin{abstract}
Within the framework of on-line learning, we study the generalization error of an ensemble
learning machine learning from a linear teacher perceptron. 
The generalization error achieved by an ensemble of linear perceptrons having homogeneous or 
inhomogeneous initial weight vectors is precisely calculated at the thermodynamic limit of a 
large number of input elements and shows rich behavior. 
Our main findings are as follows. 
For learning with homogeneous initial weight vectors, the generalization
error using an infinite number of linear student perceptrons is equal to only half that of a 
single linear perceptron, and converges with that of the infinite case with $O(1/K)$ for a 
finite number of $K$ linear perceptrons. 
For learning with inhomogeneous initial weight vectors, 
it is advantageous to use an approach of weighted averaging over the output of the 
linear perceptrons, and we show the conditions under which the optimal weights are constant 
during the learning process.  
The optimal weights depend on only correlation of the initial weight vectors.  
\end{abstract}
\maketitle
\section{Introduction}
Many ensemble learning algorithms, such as bagging \cite{Breiman1996} and the 
Ada-boost \cite{Freund1997} algorithm, try to improve upon the performance of a single 
learning machine by using many learning machines; such an approach has recently received 
considerable attention in the field of machine learning.

Theoretical analysis of the generalization error of ensemble learning has been done using 
statistical mechanics\cite{Sollich1997,Urbanczik2000}.
Sollich analyzed batch-mode ensemble learning with linear perceptrons under the noisy 
learning condition \cite{Sollich1997}, and demonstrated that the generalization error 
can be reduced by applying different subsets of entire learning examples.
Urbanczik analyzed the generalization error of ensemble learning by using simple 
perceptrons based on on-line learning\cite{Urbanczik2000}. 
He discussed two types of ensemble learning with $K$ simple perceptrons.
In the first scenario, a new simple perceptron obtained by averaging the weight vectors of 
the $K$ simple perceptrons was used as that of ensemble learning (model 1).
In the second, the average of the outputs of the $K$ simple perceptrons was used as an output 
of the ensemble learning (model 2).
Since the output property of the single perceptron is non-linear, it requires $O(e^K)$ 
calculations to theoretically obtain the generalization error for model 2 where $K$ is 
the number of simple perceptrons. 
Urbanczik mainly discussed model 1 to avoid this difficulty and demonstrated that in 
the limit of $K \rightarrow \infty$, the generalization error of model 2 converges to 
that of model 1. 

When the linear perceptron is employed, models 1 and 2 are identical, and analysis for 
a finite number of $K$ becomes possible. 
This is the main reason we employed linear perceptrons in this paper.
We assume two initial conditions to calculate the generalization error of the ensemble
learning machine: one is that the correlation between the weight vectors of learning machines 
is homogeneous, and the other is that the correlation is inhomogeneous.   
In the homogeneous case, the correlation between the weight vectors of learning machines 
will be uniform and the weight vectors will remain uniform throughout the learning process 
because of the symmetry of the evolution equation used as the update rule. 
Thus, a simple average of the $K$ outputs of the learning machine can be used as the 
ensemble output of the learning machine. 
We derived the generalization error and found that it consist of two terms; the first depends 
on the number of learning machines $K$, while the second does not depend on $K$.
It will be confirmed that the generalization error is equal to half that of a single learning 
machine when $K \rightarrow \infty$, and that the generalization error converges into that of 
the infinite case with $O(1/K)$ when $K$ is finite. 

In the inhomogeneous case, the generalization error can be improved by introducing weights 
to average the $K$ outputs of the learning machines (i.e., to obtain a weighted average rather 
than a simple average), and adapting the weights to minimize the generalization error 
(i.e., parallel boosting) \cite{lazarevic2002} is required.
We also analyze the time dependence of the weights used for averaging.
\section{Model}
\subsection{Network structure and ensemble output}
\label{size}
\begin{figure}[t]
\includegraphics[width=8cm]{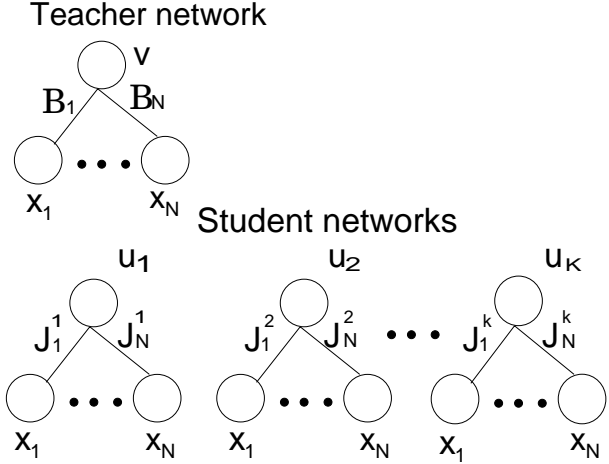}
\caption{Network structure of teacher and student networks, all having the same network structure.}
\label{network_structure}
\end{figure}

In this paper, we discuss the generalization error of ensemble learning with $K$ 
linear perceptrons. 
We assume the teacher and student networks receive $N$ dimensional input 
$\bm{x}=(x_1, \ldots , x_N)$ as shown in Fig. \ref{network_structure}, and consider the 
thermodynamic limit of $N \rightarrow \infty$.
We also assume that the elements $x_i$ of the independently drawn input $\bm{x}$ are uncorrelated 
random variables with zero mean and $1/N$ variance; that is, the elements are drawn from a 
probability distribution $P(\bm{x})$. 
The size of $\bm{x}$ is then $|\bm{x}|=1$.

\begin{equation}
\langle x_i \rangle = 0, \ \ \ \langle (x_i)^2 \rangle = \frac{1}{N}, \ \ \ |\bm{x}|=1, 
\end{equation}

\noindent
where $\langle \cdots \rangle$ means averaging over the distribution $P(\bm{x})$.

The teacher is a linear simple perceptron (as shown) that 
outputs $v(m)$ for $N$ dimensional input $\bm{x}(m) = (x_1(m), \ldots, x_N(m))$ at 
$m$th learning iterations.

\begin{eqnarray}
v(m)&=&\sum_{i=1}^N B_i x_i(m) = \bm{B} \cdot \bm{x}(m), \\
\bm{B}&=&(B_1, \ldots, B_N), 
\end{eqnarray}

\noindent
Each element $B_i$ of the teacher weight vector $\bm{B}$ is drawn from the
probability distribution of zero mean and unit variance, and is fixed throughout the learning process. 
In this case, the size of the teacher weight vector is $\sqrt{N}$, 

\begin{equation}
\langle B_i \rangle = 0, \ \ \ \langle (B_i)^2 \rangle = 1, \ \ \ |\bm{B}| = \sqrt{N}.
\end{equation}

$K$ linear simple perceptrons are used as the student networks that compose 
the ensemble learning machine. 
Each student network has the same architecture as the teacher network and 
outputs $u_k(m)$ for the $N$ dimensional input $\bm{x}(m)$.

\begin{eqnarray}
u_k(m)&=&\sum_{i=1}^N J^k_i(m) x_i(m) = \bm{J}^k(m) \cdot \bm{x}(m)\\
\bm{J}^k(m)&=&(J^k_1(m), \ldots, J^k_N(m))
\end{eqnarray}

\noindent
where $u_k(m)$ denotes $k$th student output and $\bm{J}^k(m)$ denotes weight vector of $k$th student.
The student weight vector is changed through the learning process, so the size of 
$\bm{J}^k(m)$ is assumed to be $|\bm{J}^k(m)|=l_k(m)\ \sqrt{N}$ and the size of $l_k(m)$ is $O(1)$. 
We call $l_k(m)$ the length of the student weight vector $\bm{J}^k(m)$ at $m$th learning iterations. 

The ensemble output of the student networks $\overline{u}(m)$ is given by the weighted average of  
each student network output with the weight for averaging $C_k(m)$, 

\begin{eqnarray}
\overline{u}(m)=\sum_{k=1}^K C_k(m) u_k(m) &=& \sum_{k=1}^K C_k(m) \bm{J}^k(m) \cdot \bm{x}(m), \label{output_of_student} \\
\sum_{k=1}^K C_k(m) &=& 1. \label{weight_of_student}
\end{eqnarray}

Bagging\cite{Breiman1996} is a form of ensemble learning using a fixed uniform
weight for averaging so that $C_k(m)=1/K$ throughout the learning process while parallel 
boosting\cite{lazarevic2002} uses the weighted average.

\subsection{Learning algorithm}

Learning is defined as a student network modifying the weight vector to make the output 
$u_k(m)$ approach the teacher output $v(m)$ for an given input $\bm{x}(m)$. 
We use the gradient descent algorithm to modify the student's weight vector $\bm{J}^k(m)$.
An identical input $\bm{x}(m)$ is applied to all student networks in the same order.
Therefore, all the student networks can independently learn the relation between the input
$\bm{x}(m)$ and the target $v(m)$.
The learning equation is 

\begin{equation}
\bm{J}^k(m+1) = \bm{J}^k(m)+ ( v(m)-u_k(m) ) \bm{x}(m), 
\label{learning_equation}
\end{equation}

\noindent
where $m$ denotes the iteration number.
As shown in Eq. (\ref{learning_equation}), the weight $\bm{J}^k(m)$ is
updated by using single input $\bm{x}(m)$, and then is not used again after the learning. 
This is called on-line learning.
According to the above formulation, the weight vector $\bm{J}^k(m)$ is statistically
independent of a new learning input $\bm{x}(m)$ and this makes analysis easier.

\section{Theory}

In this paper, we consider the thermodynamic limit of $N \rightarrow \infty$ to analyze 
the dynamics of the present learning system using statistical mechanics.
In the following sections, the iteration number $m$ is neglected to simplify notation of 
equations.
 
As pointed out, we will discuss learning based on on-line learning.
In on-line learning, the input $\bm{x}$ is not used after the
learning and the weight vector $J^k$ is statistically independent of a
new learning input. 
Note that the distribution of the normalized output of the student network 
$\tilde{u}_k=u_k/l_k$ obeys the Gaussian distribution of zero mean and unit 
variance at the thermodynamic limit of $N \rightarrow \infty$.
For the same reason, the distribution of teacher output $v$ obeys
the Gaussian distribution of zero mean and unit variance at the
thermodynamic limit.
Thus, the distribution $P(v,\{\tilde{u}_k\})$ of $v$ and $\{\tilde{u}_k\}$ is 

\begin{eqnarray}
&&P(v,\{\tilde{u}_k\})= \frac{1}{\sqrt{(2\pi)^{K+1}|\bm{\Sigma}|}} \exp \left (- \frac{1}{2} (v,\{\tilde{u}_k\})^{T} \bm{\Sigma}^{-1} (v,\{\tilde{u}_k\}) \right ) \label{gaussian}\\
&&\bm{\Sigma}=
   \left(
   \arraycolsep=3pt
   \begin{array}{cccccc}
     1      & R_1      & \ldots & \ldots & R_{K-1}      & R_K      \\
     R_1     & 1      & q_{1,2} & \ldots & q_{1,K-1} & q_{1,K} \\
     \vdots & q_{2,1} & \ddots &         & \vdots & \vdots \\
     \vdots &     \vdots &     &  \ddots & \vdots & \vdots  \\
     R_{K-1}  & q_{K-1,1} & \ldots  & \ldots  &    1   & q_{K-1,K}      \\
     R_K      & q_{K,1} & \ldots & \ldots  & q_{K,K-1}   & 1
   \end{array}
   \right)\label{covariance}
\end{eqnarray}

\noindent
Here, $R_k$ is the overlap between the teacher weight vector $\bm{B}$ and the
student weight vector $\bm{J}^k$ (i.e., it is the direction cosine of $\bm{B}$ 
and $\bm{J}^k$), and $q_{kk'}$ is the overlap between two student weight vectors 
$\bm{J}^k$ and $\bm{J}^{k'}$. These two overlaps, are defined as

\begin{eqnarray}
R_k &=& \frac{\bm{B} \cdot \bm{J}^k}{|\bm{B}| \cdot |\bm{J}|}=\frac{1}{Nl_k}\sum_{j=1}^N B_j J_j, \label{overlap_R}\\
q_{kk'} &=& \frac{\bm{J}^k \cdot \bm{J}^{k'}}{|\bm{J}^k| \cdot |\bm{J}^{k'}|}=\frac{1}{Nl_k l_{k'}} \sum_{j=1}^N J^k_j J^{k'}_j. \label{overlap_q}
\end{eqnarray}

\noindent
$R_k$ and $q_{kk'}$ are the order parameters of the present learning system.

\subsection{Generalization error}

The squared error of the teacher output $v$ and the ensemble output of the student 
networks $\overline{u}$ is used to evaluate the student network's performance.

\begin{equation}
\epsilon = \frac{1}{2}\left(v-\overline{u} \right)^2 = \frac{1}{2}\left ( \bm{B}\cdot \bm{x} - \sum_{k=1}^K C_k
\bm{J}^k \cdot \bm{x} \right )^2
\label{e}
\end{equation}

\noindent
Here, $C_k$ is a weight for averaging and the ensemble
output of the student networks is a weighted average of each student network's output. 
The generalization error $\epsilon_g$ is given by squared error
$\epsilon$ in Eq. (\ref{e}) averaged over the possible input $\bm{x}$ drawn from 
the Gaussian distribution $P(\bm{x})$ of zero mean and $1/N$ variance.

\begin{eqnarray}
\epsilon_g &=& \int d\bm{x} P(\bm{x}) \ \epsilon \nonumber \\
&=& \int d\bm{x} P(\bm{x}) \frac{1}{2} \left( \bm{B}\cdot \bm{x} - \sum_{k=1}^K C_k \bm{J}^k \cdot \bm{x} \right)^2
\label{generalization_error_1}
\end{eqnarray}

\noindent
For on-line learning at the thermodynamic limit, as mentioned, the distributions of the 
stochastic variable $\tilde{u}_k=(\bm{J}^k \cdot \bm{x})/l_k$ and $v=\bm{B}\cdot \bm{x}$ 
obey the Gaussian distribution of zero mean and unit variance.
Hence, the generalization error can be rewritten using Eqs. (\ref{gaussian}), 
(\ref{covariance}), 

\begin{equation}
\epsilon_g=\int dv \prod_{k=1}^K d\tilde{u}_k \ P(v, \{\tilde{u}_k\}) \frac{1}{2} \left (
v-\sum_{k=1}^K C_k u_k \right )^2.
\label{generalization_error_2}
\end{equation}

\noindent
This equation is the $(K+1)$th Gaussian integral with $\{\tilde{u}_k\}$ and $v$, and it enables 
us to make the calculation.
The result is shown by the following equation.

\begin{equation}
\epsilon_g = \frac{1}{2}\left \{ 1-2\sum_{k=1}^K C_k R_k l_k + \sum_{k=1}^K \sum_{k'=1}^K C_k C_{k'}q_{kk'}l_k l_{k'} \right\}
\label{generalization_error_3}
\end{equation}

\noindent
Consequently, the dynamics of the generalization error is calculated by
substituting the time step value of $l_k$, $R_k$, and $q_{kk'}$ into Eq. (\ref{generalization_error_3}).
Therefore, we solve the dynamics of $l_k$, $R_k$, and $q_{kk'}$ in the next subsection.
Here, $l_k, R_k$, and $q_{kk'}$ are macroscopic parameters that represent the system dynamics. 

\subsection{Dynamics of order parameters}

We first derive the dynamics of the length of student weight vector $l_k$ \cite{Nishimori2001}.
To obtain the differential equation of $l_k$, we square both sides of
equation (\ref{learning_equation}). 
We then average the term of equation (\ref{learning_equation}) by the
distribution of $P(v, \{\tilde{u}_k\})$. 
Note that $\bm{x}$ and $\bm{J}^k$ are random variables, so the
equation becomes a random recurrence formula. 
We formulate the size of the weight vectors to be $O(\sqrt{N})$, and
the size of input $\bm{x}$ is $O(1)$, so the length of student weight vector
$l_k$ has a self-averaging property.
Here, we rewrite $m$ as $m=Nt$, and represent the learning process using 
continuous time $t$. 
We obtain the deterministic differential equation of $l_k$ at the thermodynamic limit, 

\begin{equation}
\frac{dl_k}{dt}=\frac{1-l_k^2}{2l_k}
\label{differential_equation_lk}
\end{equation}

\noindent
Note that $l_k=1$ is a stable fixed point of this equation.
Next, we derive the differential equation of the overlap $R_k$ between the 
teacher weight vector $\bm{B}$ and the student weight vector $\bm{J}^k$. 
The differential equation of overlap $R_k$ is derived by calculating
the product of $\bm{B}$ and Eq. (\ref{learning_equation}), and
then averaging the term of the equation by the distribution of $P(v,\{\tilde{u}_k\})$.
The overlap $R_k$ also has a self-averaging property, and at the
thermodynamic limit, the differential equation of $R_k$ is then
obtained through a calculation similar to that used for $l_k$.

\begin{equation}
\frac{dR_k}{dt}=\frac{1}{l_k} - \frac{R_k}{2} \left (1+\frac{1}{l_k^2} \right )
\label{differential_equation_Rk}
\end{equation}

To calculate the generalization error in Eq. (\ref{generalization_error_3}) we have
to obtain the differential equation of overlap $q_{kk'}$ defined by Eq. (\ref{overlap_q}) 
\cite{Urbanczik2000}.
The overlap $q_{kk'}$ also has a self-averaging property, so we can derive the
differential equation at the thermodynamic limit. 
We define the overlap $q_{kk'}=(\bm{J}^k \cdot \bm{J}^{k'})/Nl_kl_{k'}$, the differential 
equation is derived by calculating the product of Eq. (\ref{learning_equation}) for 
$\bm{J}^k$ and the equation for $\bm{J}^{k'}$, and we obtain the 
deterministic differential equation below.

\begin{equation}
\frac{dq_{kk'}}{dt} = \frac{1}{l_kl_{k'}} - \frac{q_{kk'}}{2} \left(
  \frac{1}{l_k^2} + \frac{1}{l_{k'}^2} \right)
\label{differential_equation_qkk}
\end{equation}

\noindent
Equations (\ref{differential_equation_lk}),
(\ref{differential_equation_Rk}), and
(\ref{differential_equation_qkk}) form the simultaneous differential
equations.

\section{Results}

\subsection{Homogeneous correlation of initial weight vectors}

The component of independently drawn teacher weight vector $\bm{B}$ is as shown in Sec. \ref{size}, 
and the component of the student's weight vector $J^k_i, i=1 \sim N$ is initialized by being drawn 
from independent random variables with zero mean and unit variance in this section. 
In this case, the initial values of the order parameters are

\begin{equation}
l_k(0)=1, \ \ \ R_k(0)=0, \ \ \ q_{kk'}(0)=0.
\end{equation}

\noindent
From the symmetry of the evolution equation for updating the weight vector, 

\begin{equation}
l_k (t)\rightarrow l(t), \ \ \ R_k(t) \rightarrow R(t), \ \ \ q_{kk'}(t) \rightarrow q(t)
\end{equation}

\noindent
are obtained.
The dynamics of order parameters $l(t)$, $R(t)$, and $q(t)$ are derived as 
Eqs. (\ref{dynamics_of_l_3}), (\ref{dynamics_of_R_3}), 
and (\ref{dynamics_of_q_3}) by substituting the above conditions into 
Eqs. (\ref{differential_equation_lk}), (\ref{differential_equation_Rk}), and
(\ref{differential_equation_qkk}),

\begin{eqnarray}
\frac{dl(t)}{dt}&=&\frac{1-l(t)}{2l(t)} \label{dynamics_of_l_3}, \\
\frac{dR(t)}{dt}&=&1-R(t) \label{dynamics_of_R_3}, \\
\frac{dq(t)}{dt}&=&1-q(t) \label{dynamics_of_q_3}.
\end{eqnarray}

\noindent
Note that $l(t)=1$ is the stable fixed point of the dynamics of $l(t)$ and is given by 
solving $dl/dt=0$.

Next, we can easily solve the above equations analytically, 

\begin{eqnarray}
l(t)&=& 1, \label{dynamics_of_l_4} \\
R(t)&=&1- \exp(-t), \label{dynamics_of_R_4} \\
q(t)&=&1- \exp(-t). \label{dynamics_of_q_4} 
\end{eqnarray}

\noindent
From Eqs. (\ref{dynamics_of_R_4}) and (\ref{dynamics_of_q_4}), $q(t)=R(t)$. 
Since the order parameters do not depend on $K$, the optimal weights for averaging should 
be $C_k=1/K$.
By substituting Eqs. (\ref{dynamics_of_l_4}), (\ref{dynamics_of_R_4}), 
and (\ref{dynamics_of_q_4}) for Eq.(\ref{generalization_error_3}), the generalization 
error for $K$ student networks is rewritten as 

\begin{eqnarray}
\epsilon_g^K(t) &=& \frac{1}{2}\left ( \frac{l(t)^2(1-q(t))}{K}+(q(t)-R(t)^2)l(t)^2+(R(t)l(t)-1)^2 \right ) \\
&=& \frac{1}{2} \left \{ \frac{1-R(t)}{K} + (1-R(t)) \right \}
\label{Generalization_error_4} \\
&=& \frac{1}{2} \left \{ \frac{\exp(-t)}{K} + \exp(-t) \right \}
\label{Generalization_error_5} 
\end{eqnarray}

\noindent
$\epsilon_g^K(t)$ denotes the generalization error with $K$ student networks.
The first term on the left side of this equation depends on the
number of student networks $K$ and becomes negligible when $K$ goes to infinity.
The second term does not depend on $K$, so it remains and cannot be ignored.  
Substituting $K=1$ for Eq. (\ref{Generalization_error_4}), we 
show that the generalization error for a single student network is $\epsilon_g^1(t)=1-R(t)$, and this error  
is identical to the generalization error of a simple perceptron \cite{Nishimori2001}.
From Eq. (\ref{Generalization_error_4}), when $K$ becomes infinite, the generalization error 
of an ensemble of $K$ student networks asymptotically converges to $(1-R(t))/2$. 
Hence, the generalization error of an ensemble of $K$ student networks converges to
half that of a single student network at the limit of $K$ going to infinity. 
Note that Eqs. (\ref{Generalization_error_4}) and (\ref{Generalization_error_5})
depict dynamics of the generalization error of the ensemble because
$l(t)=1$ is the fixed point of the learning process as shown in the formulation of
the initialization given in this subsection.

\begin{figure*}[t]
\begin{minipage}[t]{7.5cm}
\includegraphics[width=7.3cm]{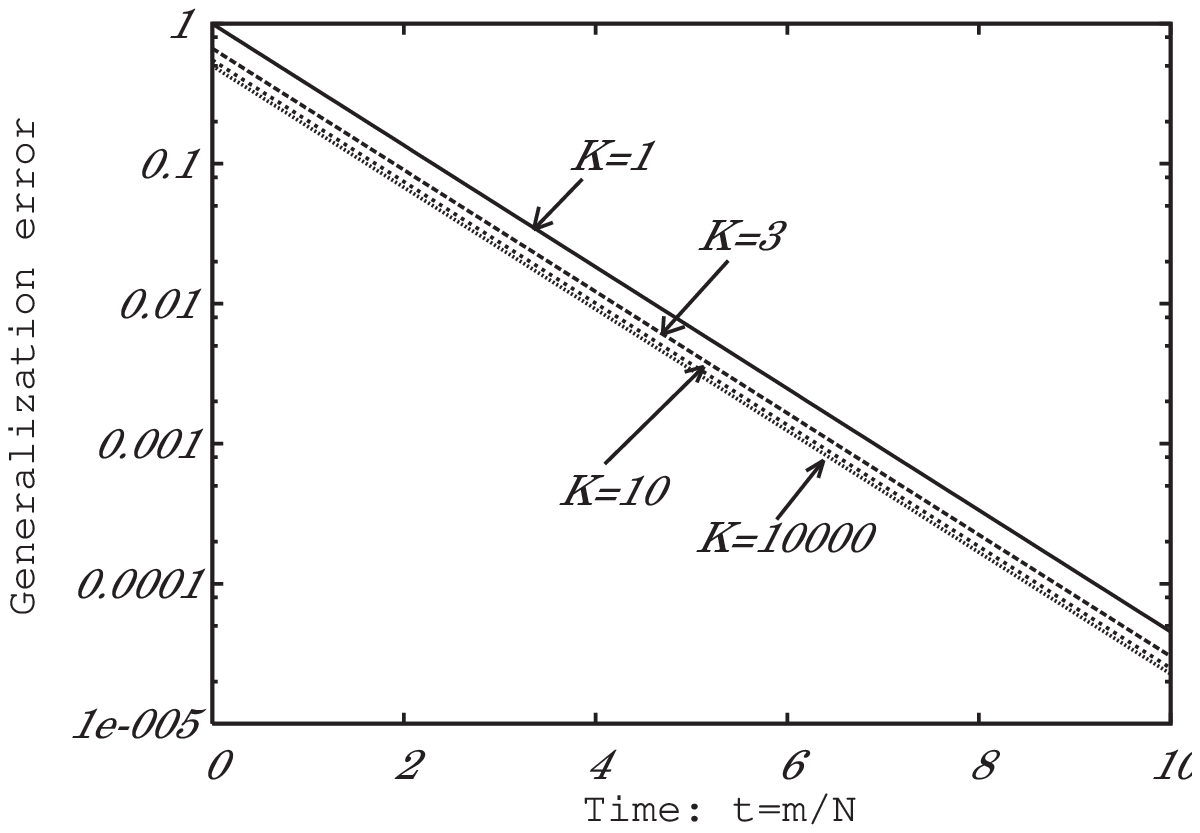}
\begin{center}
(a) Theoretical results.
\end{center}
\end{minipage}
\begin{minipage}[t]{7.5cm}

\includegraphics[width=7.3cm]{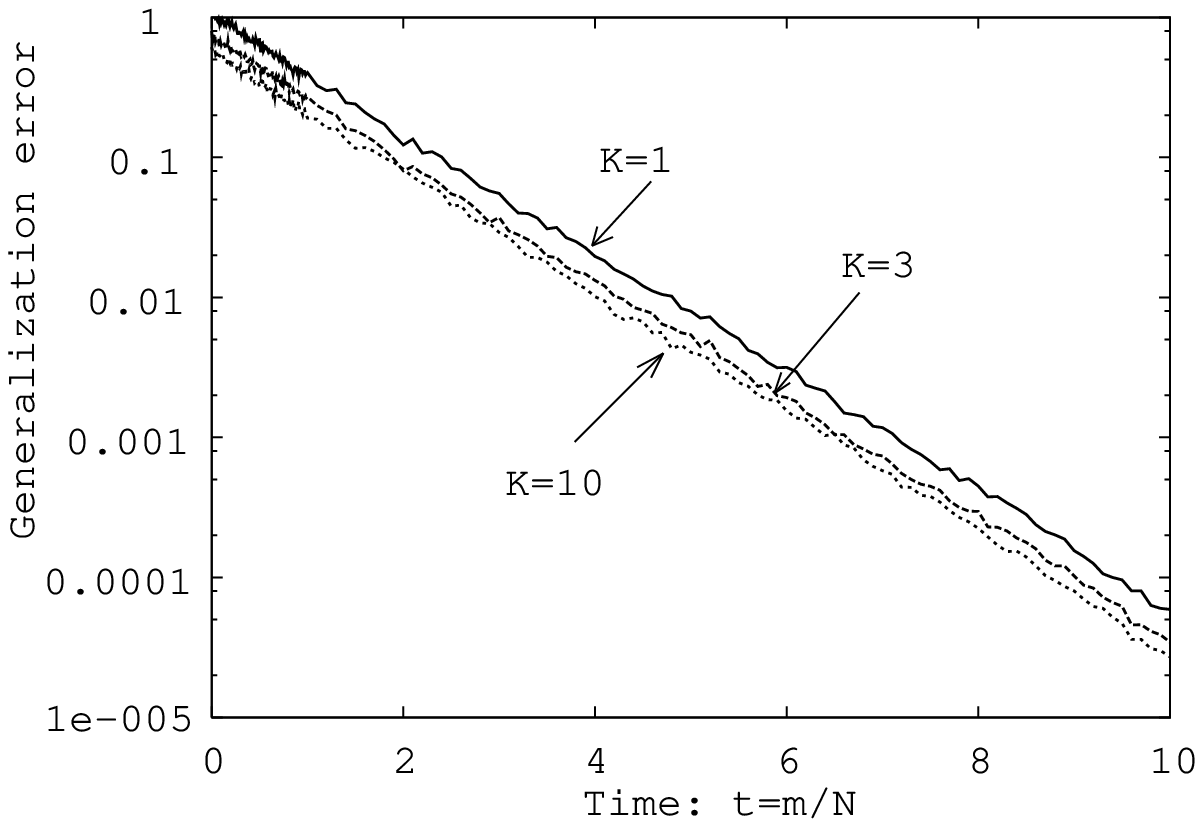}
\begin{center}
(b) Results obtained through a numerical simulation.
\end{center}
\end{minipage}
\caption{Dependence of the ensemble learning generalization error on the number of student networks $K$.}
\label{result}
\end{figure*}

Figure \ref{result} shows the $K$ dependence of the generalization error of the
ensemble: (a) shows theoretical results obtained using Eq.(\ref{Generalization_error_5}), 
and (b) shows the results obtained through a numerical simulation. 
In these figures, the horizontal axis is time $t=m/N$,
and the unit time corresponds to the time needed to feed in $N$ inputs $\bm{x}$.
The vertical axis is the generalization error $\epsilon_g$.  

First, the theoretical results for $K=1, 3, 10$, and $10000$ are shown in Fig. \ref{result}(a) 
as the diagonal lines.
The generalization error for larger $K$ ( the lower lines ) shifted and converged to half of the
generalization error for $K=1$ with respect to the order of $1/K$.
Next, the simulation results for $K=1, 3$, and $10$ are shown in Fig. \ref{result}(b).
These results agree with the theoretical results, confirming the validity of 
the theoretical results.
Hence, in the following analysis, we show only the theoretical results.

\subsection{Inhomogeneous correlation of initial weight vectors}

When the correlation between student weight vectors is inhomogeneous, for instance, if the number of 
student networks is K=3, and the initial student weight vector $\bm{J}^1(0) = \bm{J}^2(0)$ 
and $\bm{J}^3(0)$ is independent of $\bm{J}^1(0)=\bm{J}^2(0)$, it seems natural to select 
weights for averaging where $C_1=C_2=0.25$ and $C_3=0.5$, instead of using the uniform value of 
$C_1=C_2=C_3=1/3$.
This consideration suggests that the optimal generalization error can be obtained by using a weighted 
average of the student outputs when the correlation between students $q_{kk'}$ is inhomogeneous. 
This method is called ``parallel boosting''\cite{lazarevic2002}.

Because the student weight vectors are inhomogeneous, overlap $R_k$ and overlap $q_{kk'}$ 
differ from subscript $k$ or $k'$, and a weight for averaging $C_k$ will depend on 
$R_k$ and $q_{kk'}$ to reduce the generalization error.
We assume that the length of the student weight vector $l_k(0)=1$, thus it is identical for 
subscript $k$. 

The optimal weights $C_k$ satisfy the following condition,  

\begin{equation}
\frac{\partial \epsilon_g}{\partial C_k} = 0.
\label{optimal_C}
\end{equation}

Since $\epsilon_g$ is quadratic (second-order) function with respect to $C_k$, Eq. (\ref{optimal_C}) 
becomes a linear equation, and we can easily obtain the optimal $C_k$ when the order parameter $l_k(t)$, 
$R_k(t)$ and $q_{kk'}(t)$ are given. 
Order parameters $R_k(t)$ and $q_{kk'}(t)$ are time-dependent parameters, 
so the optimal weights for average $C_k$ given by Eq. (\ref{optimal_C}) generally become
time-dependent $C_k(t)$. 
In this case, we assumed $l_k(0)=1$, which then means $l_k(t)=1$ (see Eq. (\ref{dynamics_of_l_3})).
Substituting $l_k(t)=1$ and $\sum_{k=1}^K C_k(t)=1$ to Eq. (\ref{generalization_error_3}), we 
obtain the generalization error as

\begin{eqnarray}
\epsilon_g(t) &=& 1-\sum_{k=1}^{K-1}C_k(t)(R_k(t)-R_K(t))-R_K(t) \nonumber \\
&+&\sum_{k=1}^{K-1}C_k^2(t)(1-q_{kK}(t)) -\sum_{k=1}^{K-1}C_k(t)(1-q_{kK}(t)) \nonumber \\
&+&\sum_{k=1}^{K-1}\sum_{k'=2}^{K-1} C_k(t)C_{k'}(t) (1+q_{kk'}(t)-q_{kK}(t)-q_{k'K}(t)). 
\label{generalization_error_6}
\end{eqnarray}

\noindent
Equations (\ref{differential_equation_Rk}) and 
(\ref{differential_equation_qkk}) can then be solved analytically, 

\begin{eqnarray}
R_k(t) &=& 1-(1-R_k(0))\exp(-t), \label{dynamics_of_R_5}\\
q_{kk'}(t) &=& 1-(1-q_{kk'}(0))\exp(-t) \label{dynamics_of_q_5}, 
\end{eqnarray}

\noindent
where $R_k(0)$ and $q_{kk'}(0)$ are the initial values of $R_k(t)$ and $q_{kk'}(t)$, respectively.
The dynamics of the generalization error is given by substituting Eqs. (\ref{dynamics_of_R_5}) 
and (\ref{dynamics_of_q_5}).

\begin{eqnarray}
\epsilon_g(t) &=& \exp(-t) \left\{ 1-R_K(0) - \sum_{k=1}^{K-1} C_k(t)(R_k(0)-R_K(0)) \right. \nonumber \\
&+&\sum_{k=1}^{K-1}C_k^2(t)(1-q_{kK}(0))-\sum_{k=1}^{K-1}C_k(t)(1-q_{kK}(0)) \nonumber \\
&+& \left. \sum_{k=1}^{K-1}\sum_{k'=2}^{K-1}C_k(t)(1+q_{kk'}(0)-q_{kK}(0)-q_{k'K}(0)) \right \} \nonumber \\
\label{generalization_error_7}
\end{eqnarray}

\noindent
As we mentioned, the optimal weight for average $C_k$ generally depends on time $t$ because $\epsilon_g$
is a function of time $t$. 
However, in this case, $\epsilon_g$ only depends on the initial value of order parameters $R_k(0)$ and 
$q_{kk'}(0)$ as shown in Eq. (\ref{generalization_error_7}).
$\partial \epsilon_g/\partial C_k$ is also independent of time $t$.
Thus the optimal weight for average $C_k$ does not depend on time $t$ in this case.

\begin{figure}[t]
\begin{center}
\includegraphics[width=8.2cm]{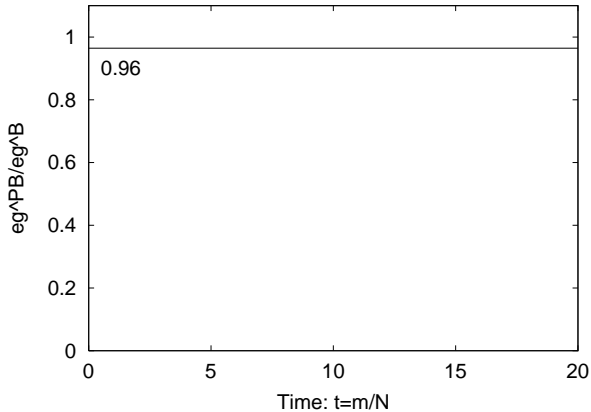}

\caption{Comparison of the generalization error with and without parallel boosting.}
\label{bagging_and_parallel_boosting}
\end{center}
\end{figure}

Figure \ref{bagging_and_parallel_boosting} shows the ratio of the generalization error with and 
without parallel boosting for $K=3$ student networks. 
Two of the student networks were identical ($\bm{J}^1(0)=\bm{J}^2(0)$), while $\bm{J}^3(0)$ was 
independent of the others. 
The generalization error using parallel boosting is denoted as \verb+eg^PB+, 
and that using bagging is denoted as \verb+eg^B+ in this figure. 
As the figure shows, parallel boosting is effective and the generalization error ratio with and 
without parallel boosting was the same throughout the learning process. 
The ratio of $\epsilon_g^{PB} / \epsilon_g^B$ was about 0.96.
 
\section{Conclusion}

We have analyzed the generalization error of an ensemble of linear perceptrons within the framework 
of on-line learning. 
Weights for averaging were introduced.
We then derived simultaneous differential equations for the order parameters $l_k$, $R_k$, 
and $q_{kk'}$ to calculate the generalization error. 
Here, $l_k$ was the length of the student weight vector $\bm{J}^k$, 
$R_k$ was the overlap between the teacher weight vector $\bm{B}$ and the
student weight vectors $\bm{J}^k$, and $q_{kk'}$ was the overlap between 
two student weight vectors $\bm{J}^k$ and $\bm{J}^{k'}$.
We have assumed two initial conditions to calculate the generalization error
of the ensemble of linear perceptrons: one was that the correlation between the weight vectors of
linear perceptrons was homogeneous, and the other was that the correlation was inhomogeneous. 

In the homogeneous case, simple averaging over the $K$ outputs of the linear 
perceptrons was valid to obtain the ensemble output of the linear perceptrons. 
We found that the generalization error was equal to half that of a single linear perceptron when 
the number of linear perceptrons $K$ became infinite, and that the generalization error 
converged into that of the infinite case with $O(1/K)$ when the number of 
linear perceptrons was finite. 

In the inhomogeneous case, the generalization error was improved by introducing
the weights for averaging over the $K$ outputs of the linear perceptrons. 
Order parameters $R_k$ and $q_{kk'}$ are time dependent, so one might think
the weight for averaging over the $K$ outputs of the linear perceptrons might be time dependent. 
However, we found the weights were not time dependent when the initial value of 
the weight length $l_k(0)=1$, and they only depended on the initial correlation between the weight 
vectors of the linear perceptrons. 
We also carried out numerical simulations whose results agreed with the theoretical results, 
thus confirming the validity of the theoretical analysis. 

\vspace{0.2cm}
\noindent
{\large \bf Acknowledgments}

\vspace{0.2cm}
Part of this study has been supported by a Grant-in-Aid for Scientific
Research in Priority Area (No. 14084212) and Grants-in-Aid for Scientific Research (C)
 (No. 13680472 and No. 14580438).
 


\begin{thebibliography}{9}
\bibitem{Breiman1996}
Breiman L., 
Machine Learning, {\bf 24}, 123 (1996).
\bibitem{Freund1997}
Freund Y., Shapire R.E.,
Journal of Comp. and Sys. Sci., {55} 119 (1997).
\bibitem{Urbanczik2000}
Urbanczik R., "Online learning with ensembles,"
Phys. Rev. E, {\bf 62}, 1448 (2000).
\bibitem{Sollich1997}
Krogh A. and Sollich P., "Statistical mechanics of ensemble learning,"
Phys. Rev. E, {\bf 55}, 1, 811 (1997).
\bibitem{Nishimori2001}
Nishimori, H., 
Statistical Physics of Spin Glasses and Information Process,
Oxford Express (2001).
\bibitem{inoue1997}
Inoue J., Nishimori H., 
Phys. Rev. E, {\bf 55}, 4, 4544 (1997).
\bibitem{lazarevic2002} 
Lazarevic A., Obradivic Z.,
Distributed and parallel databases, {\bf 11}, (2), 203, (2002).
\end{thebibliography}
\end{document}